# Focal shift of silicon microlens in mid-infrared regime

Haijie Zuo,[a] Jiangyong Zhang,[a] Leiying Ying [a], Baoping Zhang,[a*], Zhijin Hou, [b] Hongxu Chen, [b] Junjie Si[b]

[a] Xiamen University, Laboratory of Micro/Nano Optoelectronics, Department of Electronic Engineering, Xiamen, China, 361005
[b] Luoyang Optoelectro Technology Development Center, Luoyang, Henan, China, 471099

**Abstract**. In this study, rigorous numerical calculation was utilized to characterize the focal properties of mid-infrared silicon microlens with the size about tens of micrometers. It is found that the focal shift phenomenon also exists in mid-infrared regime, which behaves differently from that of visual and near-infrared wavelength. Focal properties of silicon microlens were also measured experimentally, showing well coherence with simulation results. Our results provide systemic understanding of focal shift in mid-infrared regime, at that wavelength special consideration should be paid in micro-nano optics, especially with the integration between infrared optical system and other devices.

*Baoping Zhang**, E-mail: bzhang@xmu.edu.cn

## 1 Introduction

With the popularization of micro-nano fabrication techniques, such as lithography and induced coupled plasma (ICP) etching, various new kinds of micro-nano optical components are being fabricated[1,2] and integrated in many applications ranging from imaging to light concentration, such as solid state image sensors[3-5], detection[6], solar cells and LEDs[7]. Being compatible with standard semiconductor fabrication process[8], physical dimension of these devices reaches the range of several wavelengths. Miniaturization of these devices results in deterioration of focal properties derived from traditional and rigorous analysis model, among which one of the widely known phenomena is focal shift[9], namely the difference between actual focal length and designed one.

Over the decades, extensive studies have been conducted to explore the properties of focal shift and its effect on the integration in optical system[3, 10-14]. Most studies, however, were focused in the visual or near-infrared wavelength band, few literature studied this phenomena in mid-infrared regime, in which silicon microlens[15, 16], for example, are usually employed in image sensor system because of its high optical transmittance and stable thermal and mechanical properties. Classical diffraction theory was usually adopted to perform the calculation of focal shift in previous studies, however, it is known that traditional analysis models should be precluded when the dimension of optical devices reaches to the order or smaller than the illumination wavelength. Thus, in order to get a more accurate understanding of focal shift in wavelength scale, rigorous numerical method (3D FDTD) was employed to study the focal shift phenomena of silicon microlens in mid-infrared regime. Based on our simulation, silicon microlens was then fabricated and corresponding measurement was carried out to characterize focal shift properties, which is to our knowledge the first time that focal shift was founded and studied experimentally in mid-infrared regime.

## 2  Numerical calculation

Traditional analysis models employing scalar numerical calculation were usually used to analysis the properties of microlens, such as Rayleigh-Sommerfeld (R-S) integral[17, 18], boundary diffraction wave theory[19], Gaussian beam decomposition algorithm[20], etc. Deviation between classical theory and rigorous vector theory in predicting focal properties become obvious with the miniaturization of device dimension[21, 22], hence vector analysis models should be utilized. Among other rigorous calculation method[23-26], we hereby employed full 3D FDTD package[27, 28] to model the silicon microlens because its time efficiency and compatibility with many complex structures.

The structure of the simulated multistep microlen was obtained based on optical path difference (OPD) using traditional scalar method[29]. In order to explore focal shift in mid-infrared wavelength band, 4.2 μm, which is among the center of mid-infrared atmospheric window, was chosen as the illumination wavelength in simulation setup.

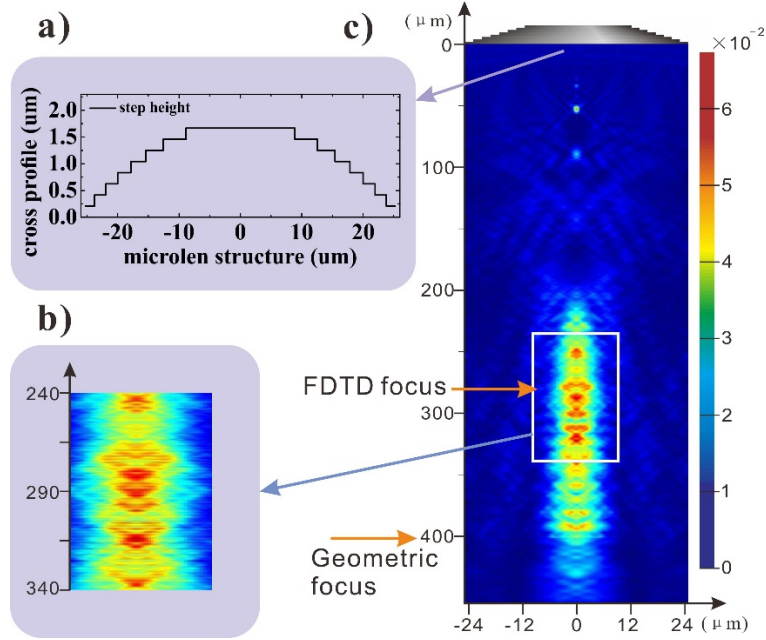

**Fig. 1** Schematic simulation setup and power distribution of silicon microlen. (a) Geometry of multistep microlen consisting 8 discrete steps with each step height of about 0.2μm, size of microlens size is 50*50*1.7(μm), and Si substrate thickness is 400μm. (b) Power distribution along wave illumination direction (z-axis), incident plane center wavelength is 4.2μm and bandwidth is 0.5μm. (c) The inset shows the magnified details of focal properties.

The basic simulation setup is shown in Fig. 1, mid-infrared plane wave incident vertically upon the multistep surface of silicon surface, curved phase front caused by these steps lead to focusing effect within the Si substrate. As can be seen in Fig. 1, the point of absolute power intensity (FDTD focus, known as actual focal spot) does not coincide with the geometric focal spot (Geometric focus), but shifted along z-axis toward the microlen surface. A focal depth of about 80 μm can also be found for this microlen with a low Fresnel number N.

Several papers tried to address the focal shift theoretically for microlens in low-Fresnel-number systems[19, 30-32], where focal shift was defined as $\Delta f = f - f'$, f is geometric focal length, and f′ is FDTD focal length. These theories, which derived from visible wavelength, failed to predict the focal shift in this simulation setup, because of the incident wavelength here is mid-infrared. Formulas in these papers suggested a relative focal shift $(\Delta f/f)$ of about 0.26~0.28 for this structure (Fresnel number N=1.49, defined by $= \alpha^2/\lambda f$, f number $\gamma=4$, defined by $\gamma = f/2\alpha$, where $\gamma$ and $\lambda$ are the radius of microlens and operational wavelength, respectively), whereas the value obtained by 3D FDTD is about 0.25~0.35, as shown in Fig. 1. We attribute this mismatch to the limited validity of classical theory: as the decrease of physical dimension of microlens to wavelength scale and extension of wavelength band to mid-infrared region, rigorous analysis method has to be utilized to more accurately model optical properties of these kinds of optical devices.

## 3   Fabrication and characterization

The Si microlens were successfully fabricated, requiring only triple lithography alignment with each followed by induced coupled plasma (ICP) etching[8]. The etching depth, however, should be carefully controlled in order to construct appropriate focusing phase front. We also calibrated the etching parameters to get a smooth surface after etching. As we can see in Fig. 2, steps obtained by etching show high fidelity to the design structure.

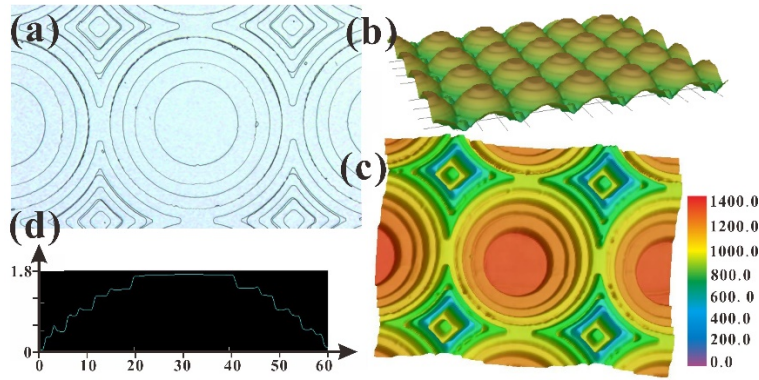

**Fig. 2** 3D laser microscope characterization of silicon microlens. (a) High resolution optical photograph, top view. (b-c) Bird's eye-view and top view of silicon microlens. (d) Measured cross section profile of silicon microlens.

AFM measurement shows that surface RMS increase with the etching depth, the largest RMS in 8th step is lower than 3nm, indicating that surface scattering can be neglected in our simulation[33].

Then ZnS thin film antireflective coating was then conducted on the bare surface of 400 um thick silicon wafer using RF magnetron sputtering to exclude the reflective power loss.

## 4    Focal properties measurement

*4.1 Experimental setup*

Focal properties of microlens was experimen1tally studied on a horizontal "microscope" system. Silicon surface with microlens was illuminated using collimated plane mid-infrared wave from standard blackbody light-source, the focal spot behind the microlens was imaged with a high magnification on to an HgCdTe camera (InfraTec, ImageIR 8800 Series). As is shown in Fig. 3, firstly, the positions of microlen array and camera were fixed, we adjusted the location of objective along optical axis, by doing this we set the focal plane of objective at the surface of silicon microlens array, this can be confirmed when clear image of the structure of concentric circles on silicon surface can be observed by camera. As we know it's hard to obtain high resolution image of microlens structure details on the order of one or two μm using one set of objective lens, especially with the illumination of long mid-infrared incident wave. Then we adjust the position of microlens using 3D step stage by displacing microlens array toward z positive direction. As the silicon surface was illuminated by plane wave, diffraction pattern within the silicon substrate (z<0) doesn't change with respect to the microlens surface. As a result, a series of diffraction patterns at different location within the silicon substrate could be captured with the displacement of microlens array.

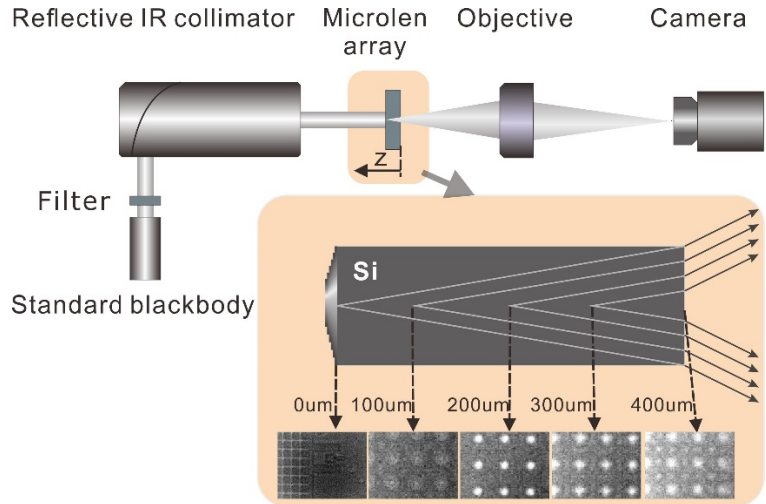

**Fig. 3** Experiment setup of focal properties measurement. Mid-infrared plane wave incident vertically in microlen array, focusing properties behind the microlen can be monitored by camera through objective as depicted in the inset.

As is shown in Fig. 3, when we adjusted the position of microlens along optical axis, different diffraction patterns within the microlens substrate can be obtained. The position of image plane was recalculated to account for refraction of imaging light at the output Si/air surface, thus diffraction pattern of focusing light can be mapped along z-axis. One thing that has to be pointed out was that the image obtained using HgCdTe camera was generated by digital voltage of incident signal, which needs to be transformed to the equivalent blackbody temperature, and then we can get the power distribution of each picture by integral of blackbody radiation formula at these temperatures.

*4.2 Measurement results*

Focal power intensity ( I(z), as is shown in Eq. 1, d and power pattern E is shown in inset of Fig. 4) at 8 discrete z-vertical planes after the silicon was plotted along the incident direction in Fig. 4, as well as the power intensity along optical axis calculated using full electromagnetic field simulations in Part 2. Power patterns at planes at different planes of consecutive z values was captured: from the plane of microlen surface back to the plane located outside of silicon wafer with a distance of 200 μm (measuring rang is 0~600 μm).

$$I(z) = \int_0^{d/2} E\, ds, \qquad (1)$$

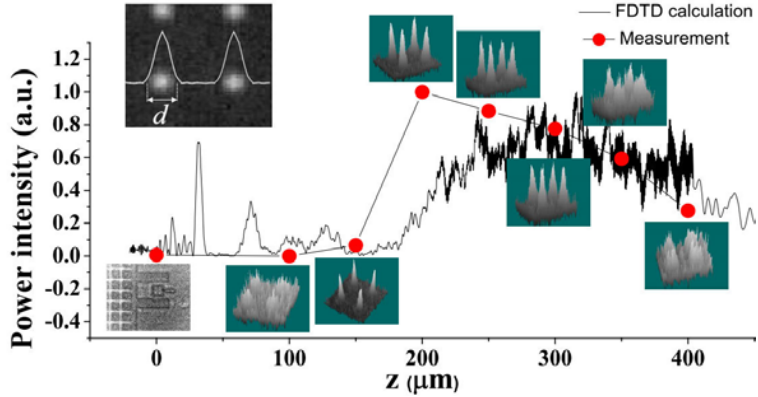

**Fig.4** Maximum power concentration of focusing light along optical axis of microlens.

Considering the measurement error caused by the limitation of objective lens and long incident wavelength, our measurement results show good agreement with electromagnetic simulation. For silicon microlens in this paper with Fresnel number of 1.49 and f-number of 4, the focal spot (location of maximum power intensity) is located at around 260~300 μm with a focal depth of about 40 μm, where its geometric focal length was design at 400 μm, indicating a focal shift with a factor of 0.25~0.35, considering the rather dispersed power spot, which was referred as depth of focal.

## 5 Conclusions

In conclusion, focal shift of silicon microlens in mid-infrared regime was theoretically proposed and experimentally verified. We reviewed the traditional method in modeling the focal shift phenomena in visual or near-infrared wavelength band, it is founded that these method reach their limits of validity for system of small structure and long wavelength band in mid-infrared. Rigorous electromagnetic method (3D FDTD) was then employed to model the focal shift effect of a silicon microlens in mid-infrared regime. Numerical calculation reveals the existence of focal shift with a factor of about 0.25~0.35. We have also traced focal shift of silicon microlens in mid-infrared regime experimentally. The rigorous calculation method in this paper provides reference value when dealing with focal shift effect, as well as other optical properties, in micro-nano optical devices with critical dimension on the order of wavelength scale. Moreover, the demonstration of focal properties measurement also provides ways to characterize the focusing effect experimentally.

Our results are of practical value in the integration of micro-nano optical systems, such as: silicon microlens in focal plane array mid-infrared photodetector, where distortion in predicting the focal length of silicon microlens between traditional theory and rigorous model becomes significant with the miniaturization of device size. In these cases the focal length of microlens has to be confirmed using rigorous electromagnetic method and measured experimentally, in this way the focal spot of incident power will thus be located exactly at the photosensitive area of photodetector.

*Acknowledgments*

This work was supported by the Special Project on the Integration of Industry, Education and Research of Aviation Industry Corporation of China.